\documentclass[12pt,preprint]{aastex}
\def\degpoint{\ifmmode ^{\rm{o}}\!. \else $^{\rm{o}}\!.$\fi}

\newcommand{\ms}{\mbox{m\,s$^{-1}$}}

\newcommand{\Msun}{\mbox{M$_{\odot}$}}
\newcommand{\Rsun}{\mbox{R$_{\odot}$}}
\newcommand{\Mjup}{\mbox{M$_{\rm Jup}$}}

\newcommand{\ltsimeq}{\raisebox{-0.6ex}{$\,\stackrel
         {\raisebox{-.2ex}{$\textstyle <$}}{\sim}\,$}}

\begin{document}

\title{The Weihai Observatory search for close-in planets orbiting giant stars }

\author{Robert A.~Wittenmyer\altaffilmark{1}, Dongyang Gao\altaffilmark{2}, Shao 
Ming Hu\altaffilmark{2}, Eva Villaver\altaffilmark{3}, Michael 
Endl\altaffilmark{4}, Duncan Wright\altaffilmark{1} }

\altaffiltext{1}{School of Physics and Australian Centre for Astrobiology, 
Faculty of Science, The University of New South Wales, Sydney 2052, Australia}
\altaffiltext{2}{Shandong Provincial Key Laboratory of Optical Astronomy and 
Solar-Terrestrial Environment, Institute of Space Sciences, Shandong University, 
Weihai, 264209, China}
\altaffiltext{3}{Departmento F\'isica Te\'orica, Facultad de Ciencias, 
Universidad Aut\'onoma de Madrid, Cantoblanco, 28049, Madrid, Espa\~na}
\altaffiltext{4}{McDonald Observatory, University of Texas at Austin, 1 
University Station C1400, Austin, TX 78712, USA}

\email{
rob@unsw.edu.au}

%\shorttitle{Weihai Planet Search }
\shortauthors{Wittenmyer et al.}

%-------------------------------------------------------------------
\begin{abstract}

Planets are known to orbit giant stars, yet there is a shortage of 
planets orbiting within $\sim$0.5 AU ($P\ltsimeq$100 days).  
First-ascent giants have not expanded enough to engulf such planets, but 
tidal forces can bring planets to the surface of the star far beyond the 
stellar radius.  So the question remains: are tidal forces strong enough 
in these stars to engulf all the missing planets?  We describe a 
high-cadence observational program to obtain precise radial velocities 
of bright giants from Weihai Observatory of Shandong University.  We 
present data on the planet host Beta Gem (HD\,62509), confirming our 
ability to derive accurate and precise velocities; our data achieve an 
rms of 7.3\,\ms\ about the Keplerian orbit fit.  This planet-search 
programme currently receives $\sim$100 nights per year, allowing us to 
aggressively pursue short-period planets to determine whether they are 
truly absent.

\end{abstract} 


%--------------------------------------------------------------------
\section{Introduction}

Searches for planets orbiting evolved stars have been underway for more 
than a decade.  A major science goal of these campaigns has been to 
explore the dependence of planetary system properties on host star mass 
-- precise radial velocities can be obtained for these ``retired A 
stars'' whereas A-type dwarfs present few absorption lines from which 
Doppler information can be extracted.  However, of the 94 planets known 
to orbit giant stars (log $g < 3.5$), only seven have semimajor axes 
$a<0.5$\,AU.  This is in spite of a strong bias favoring the detection 
of short-period planets. About 2000 evolved high-mass stars are 
currently being monitored worldwide by various teams, 
\citep[][e.g.]{hatzes93, frink02, sato03, setiawan03, hatzes05, 
johnson06, n08, lee11, 47205paper}.  Given that short-period planets are 
common around solar-mass stars, there is an open question: Where are the 
short-period planets around high-mass stars?

\citet{vl09} proposed that planetary orbits are affected by the 
evolution of the stars, showing how tidal interaction can lead to the 
engulfment of close-in planets.  This process is strongly influenced by 
stellar evolution details such as the stellar mass, mass-loss, and 
metallicity.  Engulfment appears to be much more efficient for 
more-massive planets and less-massive stars \citep{kunitomo11, 
villaver14}.

Giant and subgiant stars sample planet hosts that, in principle, are 
more massive than their main sequence counterparts 
\citep[e.g.][]{johnson10, sato08, bowler10, gettel12}.  This supposition 
(taken alone) has been used to suggest a relation between planet 
formation and stellar mass \citep{currie09}.  We note that the masses of 
these evolved stars have been the subject of some controversy, e.g. 
\citet{lloyd11, lloyd13, johnson13}.  In brief, \citet{lloyd11} argues 
that due to the stellar initial mass function and the rapid evolution of 
more massive stars, a given evolved star is more likely to be a 
``retired'' Sun-like star than a ``retired A star'' as put forward by 
\citet{johnson06} and \citet{johnson13}.  While this relation might 
still hold, there is, however, an increasing number of close-in planets 
being detected in the main-sequence stage orbiting A-F stars (e.g., 
HAT-P-49, Bieryla et al. 2014; WTS-1b, Cappetta et al. 2012; Kepler-14b, 
Buchhave et al. 2011; WASP-33b, Collier Cameron et al. 2010; KELT-3b, 
Pepper et al. 2013; OGLE2-TR-L9b, Snellen et al. 2009).  These close-in 
planets are not found in radial velocity searches around evolved stars, 
suggesting that it is the evolution of the star, and not its mass, that 
plays a major role in removing planets from close orbits.  Thus the 
discovery of a population of short period planets around evolved stars 
can offer fundamental insights into the strength of the tidal forces.  
Understanding these tidal forces is critical for producing accurate 
models of planetary orbits as stars evolve off the main sequence.  
Furthermore, they can help to shed some light on the dependency of the 
planet formation process on the stellar mass, with important 
implications for gas giant planet formation and migration.

In fact, close-in planets are expected to enter the stellar envelope as 
the star leaves the main sequence and evolves onto the red giant branch 
\citep{vl09, villaver14}.  As the star evolves, it removes planets from 
a region that extends far beyond the stellar radius to the entire region 
of tidal influence ($a/R_{*}\sim\,2-3$ for a Jupiter-mass planet).  
During the subgiant phase, the star first clears out the very close-in 
planets present during the main-sequence evolution, and then proceeds to 
clear out a larger region as the stellar radius increases when it 
ascends the red-giant branch.  However, the vast majority of the evolved 
stars targeted by radial-velocity surveys have not completed their 
ascent up the red-giant branch, and so are not at their maximum radius.  
The evolved stars currently known to host planets have typical radii 
smaller than about 6\Rsun\ \citep{sato10}.  Stellar evolution models by 
\citet{vl09} show that for a 2\Msun\ star, the radius exceeds 0.1\,AU 
(i.e. a 10-day orbital period) for only the 40 million years immediately 
preceding the helium flash.  For these reasons, we only expect planets 
orbiting closer than $a\,\sim$0.15 AU to be lost to their expanding host 
star, and this only excludes planets with $P\ltsimeq$15 days 
\citep{johnson07}.  Our survey will search for the planets with $15 < P 
< 100$ days which are common around main-sequence stars, but missing 
around evolved stars.

This paper is organised as follows: Section 2 describes the Weihai 
Echelle Spectrograph (WES), and the target selection criteria.  Section 
3 demonstrates the ability of the WES to deliver precise radial 
velocities by presenting new data and orbital fits for the known 
planet-hosting giant Beta Gem (HD\,62509).  Finally, we give our 
conclusions in Section 4.

%----------------------------------------------------------
\section{Observational Program}

\subsection{The Weihai Observatory Echelle Spectrograph}

The Weihai Echelle Spectrograph (WES) is a bench-mounted, stabilised, 
fibre-fed spectrograph attached to the 1-metre telescope located at the 
Weihai Observatory of Shandong University in Weihai, China.  The 
spectrograph is thermally stabilised to $\pm$0.10 K.  Light is fed from 
the telescope into the spectrograph by $\sim$10m of circular fibre with 
diameter 70$\mu$m via a 160$\mu$m pinhole.  It can achieve a maximum 
resolution of 57,000 with 2.2-pixel sampling.  The WES is the first 
fibre-fed echelle spectrograph in China, and has the primary function of 
radial-velocity planet search.  Further details about the spectrograph 
and the Weihai Observatory site can be found in \citet{gao14}, 
\citet{hu14}, and \citet{guo14}.

The observing and data analysis procedures are typical of precise 
radial-velocity planet searches.  Doppler velocity observations are 
performed at a resolving power of $R\sim$45,000.  Calibration of the 
spectrograph point-spread function is achieved using an iodine 
absorption cell temperature-controlled at 65.0$\pm$0.1$^{\rm{o}}$C.  The 
iodine cell imprints a dense forest of narrow absorption lines on the 
stellar spectrum, from 5000 to 6200\,\AA, permitting the contemporaneous 
calibration of the spectrograph point-spread function 
\citep{val:95,BuMaWi96}.  Velocities are obtained using the 
\textit{Austral} code \citep{endl00}, which has been successfully used 
by several planet-search programs for more than 10 years 
\citep{endl04,47205paper,texas1}.  The iodine region is broken into 
$\sim$380 100-pixel chunks, corresponding to about 4.3\,\AA per chunk.  
The chunks are weighted by their Doppler content (defined as the sum of 
all pixel to pixel gradients), and each chunk produces a radial velocity 
relative to the iodine-free template spectrum.  The final velocity is 
computed as the mean value of the chunks after an iterative 3$\sigma$ 
clipping to reject outliers.  The uncertainty is the standard error of 
the mean, i.e. the rms scatter divided by the square root of the number 
of accepted chunks.

\subsection{Target selection and observing strategy}

We chose a small sample of bright giants, starting with Northern 
hemisphere \textit{Hipparcos} stars with $V<5.0$ and luminosity classes 
III and IV \citep{vl07}.  After removing all stars flagged for 
variability, double/multiple stars, and suspected composite spectra, we 
imposed a colour cut $0.5<(B-V)<1.2$ to match the selection criteria of 
many evolved-star planet-search surveys \citep[e.g.][]{sato05, 
johnson06, n08, jones11}.  We then required $M_V > 0.0$ to exclude the 
most-evolved stars which would be subject to large-scale pulsations that 
confound radial-velocity searches for planets \citep{hekker08}.  To 
reduce the impact of pulsations on the detectability of planets, we also 
required that the \textit{Hipparcos} photometric scatter be smaller than 
0.005 mag.  From the scaling relations of \citet{kb95}, this threshold 
corresponds to velocity amplitudes smaller than 5-10\,\ms, a level 
comparable to our photon noise and small compared to the expected 
amplitudes of giant planets in short-period orbits.  The 42 remaining 
targets are enumerated in Table~\ref{targetlist}.

Our knowledge of short-period planets orbiting these evolved stars 
remains limited by the observing strategies employed by the major 
programs.  Radial-velocity surveys are usually subject to the exigencies 
of telescope scheduling, such that they are typically allocated time in 
a single short run each month (during the bright lunation).  The ``Rocky 
Planet Search'' campaigns conducted by the Anglo-Australian Planet 
Search addressed this problem by observing a subset of $\sim$30 stars 
every night for 48 consecutive nights.  Those campaigns have 
demonstrated dramatically increased sensitivity to short-period planets 
\citep{16417paper,vogt10,etaearth}.  Clearly, high cadence is the way 
forward \citep{swift14}.  For example, \citet{jones15} were able to 
conclusively detect the 89-day planet orbiting the K giant HD\,121056 by 
densely sampling its orbit using CHIRON, a spectrograph dedicated to 
precise radial-velocity planet search \citep{chiron}.  To achieve the 
best possible cadence, our survey has been allocated $\sim$100 nights 
per year on the Weihai Observatory 1m telescope.

%----------------------------------------------------------------
\section{Preliminary Results: Beta Gem's Planetary Companion}

Six of our 42 targets are known planet hosts (Table~\ref{planethosts}).  
They are useful members of the sample because (1) further monitoring can 
often reveal additional planets \citep{wright09, songhu}, and (2) the 
repository of data on these systems can serve as a check on the new data 
being obtained by this survey.

Beta Gem (HD\,62509, HIP\,37826) is an extremely bright K0 giant 
($V=1.14$) known to host a planet with $P=590$ days and m~sin~$i$=2.76 
\Mjup\ \citep{hatzes93,hatzes06}.  We have 38 observations of HD\,62509 
over a 500-day baseline, with a mean internal velocity uncertainty of 
8.6\ms.  Exposure times ranged from 300-900 seconds, with typical 
S/N$\sim$200-300 per pixel.  An iodine-free template spectrum was 
obtained on 2014 March 7, and all velocities given in 
Table~\ref{weihaivels} are computed relative to that template.

To check the consistency of the Weihai data, we include all published 
velocities spanning more than 25 years to fit a Keplerian orbit to the 
planetary signal.  Our analysis uses eight data sets, seven of which 
have already been published.  \citet{larson93} presented velocities from 
the CFHT ($N=39$) and DAO ($N=27$), calibrated using an HF gas cell 
\citep{campbell79}, a precursor to the currently-used iodine cell 
technique.  The Beta Gem planet discovery paper \citep{hatzes93} gave 38 
radial-velocity measurements from the McDonald Observatory 2.1m 
telescope.  \citet{hatzes06} reported a further 22 epochs from the 
ongoing McDonald Observatory 2.7m ``Phase 3'' planet-search 
\citep[e.g.][]{texas1,texas2} and 11 epochs using a higher-resolution, 
limited wavelength setting on the same telescope (``cs21'' as desribed 
in Hatzes et al. 2006 and Wittenmyer et al. 2006).  The Tautenburg 
Observatory Planet Search (TOPS) also yielded 22 radial velocities as 
presented in \citet{hatzes06}.  Finally, 80 velocities from Lick 
Observatory given in \citet{reffert06} were included in our fit.

%  2coude = cs21 on 2.7m
%  phase 1 = HC93 paper, 2.1m  

\begin{figure}
\plotone{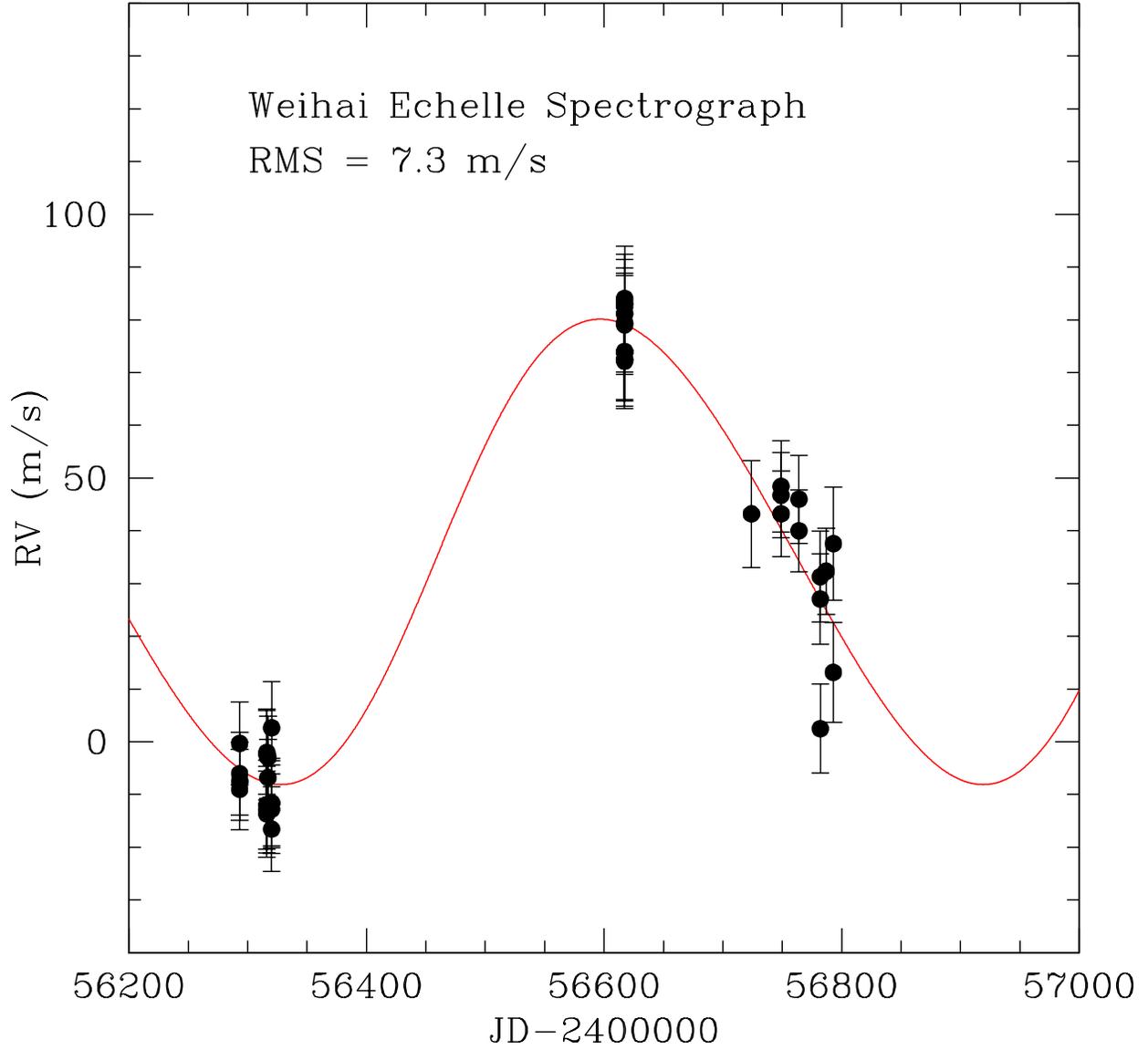}
\caption{WES data for Beta Gem: 38 observations from 2013 Dec 31 to 2014 
May 15.  Our results are in excellent agreement with the published data, 
now spanning 33.5 years and confirming the consistency of the planet's 
orbital parameters. }
\label{plot}
\end{figure}

We used the \textit{GaussFit} nonlinear least-squares code 
\citep{jefferys87} to obtain a Keplerian model fit for all eight data 
sets simultaneously.  The velocity data now span 33.5 years.  
Uncertainties were estimated using the bootstrap routine within 
\textit{Systemic 2} \citep{mes09} on 10,000 synthetic data set 
realisations.  The results are given in Table~\ref{planetparams}, and 
the most recent cycles including the Weihai data are plotted in 
Figure~\ref{plot}.  The rms about the fit for the Weihai velocities is 
7.3\,\ms, comparable to the previously published data, and consistent 
with the stellar oscillation amplitude of 5-6\,\ms\ found by 
\citet{hatzes12}.  This result demonstrates that the Weihai data 
acquisition and Doppler velocity extraction techniques are robust.

%----------------------------------------------------------
\section{Summary and Conclusions}

Close-in planets ($a<0.5$ AU, $P\ltsimeq$100 days) are rare around 
giant-branch stars.  We have begun an observational program at the 1m 
telescope located at Weihai Obervatory of Shandong University, using its 
stabilised echelle spectrograph to obtain precise Doppler velocity 
measurements of a sample of bright giant stars.  We aim to observe these 
stars with as high a cadence as possible to search for these ``missing'' 
planets.  Preliminary results from this progam give excellent results 
for the known planet-host Beta Gem (HD\,62509), and confirm our ability 
to obtain precise velocities with this new instrument.

%----------------------------------------------------------
\acknowledgements

The Weihai iodine cell is provided by Bun'ei Sato and the Okayama 
Astrophysical Observatory.  RW acknowledges support from UNSW Faculty 
Research Grants.  S. M. Hu would like to thank the support by the 
National Natural Science Foundation of China and Chinese Academy of 
Sciences joint fund on astronomy under grant No. U1331102, by the 
National Natural Science Foundation of China under grant No. 11333002 
and by Sino-German Science foundation under project No. GZ788.  M.E. is 
supported by the National Science Foundation through grant AST-1313075.

%--------------------------------------------------------------------

%-------------------------------------------------------------------

%  ADD FE/H, TEFF, LOGG FOR ALL STARS

\begin{deluxetable}{lcccccccc}
\tabletypesize{\scriptsize}
\tablecolumns{9}
\tablewidth{0pt}
\tablecaption{Weihai target list }
\tablehead{
\colhead{HIP} & \colhead{HD} & \colhead{RA} & \colhead{Dec} & \colhead{$V$} & 
\colhead{$T_{eff}$ (K)} & \colhead{log $g$ (cgs)} & \colhead{[Fe/H]} & 
\colhead{Reference}
}
\startdata
\label{targetlist}
        3031 & 3546   &  00 38 33.50  & +29 18 44.5  &  4.34 & 5102 & 2.80 & -0.56 & \citet{luck07} \\
        4422 & 5395   &  00 56 40.01  & +59 10 52.2  &  4.62 & 4875 & 2.7 & -0.51 & \citet{mass08} \\
        4906 & 6186   &  01 02 56.66  & +07 53 24.3  &  4.27 & 4955 & 2.68 & -0.24 & \citet{luck07} \\
        5586 & 7106   &  01 11 39.59  & +30 05 23.0  &  4.51 & 4748 & 2.62 & 0.01 & \citet{luck07} \\
        6411 & 8207   &  01 22 20.39  & +45 31 43.5  &  4.87 & 4656 & 2.8 & 0.03 & \citet{mass08} \\
        7294 & 9408   &  01 33 55.93  & +59 13 55.5  &  4.68 & 4900 & 2.67 & -0.22 & \citet{luck07} \\
        9884 & 12929   &  02 07 10.29  & +23 27 46.0  &  2.01 & 4498 & 2.4 & -0.25 & \citet{mass08} \\
       13061 & 17361   &  02 47 54.44  & +29 14 50.7  &  4.52 & 4727 & 2.67 & 0.07 & \citet{luck07} \\
       14668 & 19476  &  03 09 29.63  & +44 51 28.4  &  3.79 & 5022 & 3.12 & 0.19 & \citet{luck07} \\
       14838 & 19787   &  03 11 37.67  & +19 43 36.1  &  4.35 & 4732 & 2.7 & -0.03 & \citet{mass08} \\
       19038 & 25604   &  04 04 41.66  & +22 04 55.4  &  4.36 & 4699 & 2.6 & 0.01 & \citet{mass08} \\
       20252 & 27348   &  04 20 24.66  & +34 34 00.3  &  4.93 & 5073 & 3.10 & 0.09 & \citet{luck07} \\
       20455 & 27697   &  04 22 56.03  & +17 32 33.3  &  3.77 & 5058 & 3.02 & 0.19 & \citet{luck07} \\
       20877 & 28292   &  04 28 26.37  & +16 21 34.7  &  4.96 & 4529 & 2.5 & -0.17 & \citet{mass08} \\
       20885 & 28307   &  04 28 34.43  & +15 57 44.0  &  3.84 & 4955 & 2.9 & 0.04 & \citet{mass08} \\
       20889 & 28305   &  04 28 36.93  & +19 10 49.9  &  3.53 & 4797 & 2.6 & 0.04 & \citet{mass08} \\
       22957 & 31421   &  04 56 22.32  & +13 30 52.5  &  4.06 & 4498 & 2.4 & -0.26 & \citet{mass08} \\
       24822 & 34559   &  05 19 16.59  & +22 05 48.1  &  4.96 & 5096 & 3.17 & 0.10 & \citet{luck07} \\
       26366 & 37160   &  05 36 54.33  & +09 17 29.1  &  4.09 & 4898 & 2.9 & -0.63 & \citet{mass08} \\
       26885 & 37984   &  05 42 28.66  & +01 28 28.8  &  4.90 & 4508 & 2.2 & -0.55 & \citet{mass08} \\
       27483 & 38656   &  05 49 10.46  & +39 10 52.1  &  4.51 & 5021 & 2.90 & -0.09 & \citet{luck07} \\
       28358 & 40035   &  05 59 31.55  & +54 17 05.9  &  3.72 & 4911 & 2.86 & -0.01 & \citet{luck07} \\
       29696 & 43039   &  06 15 22.74  & +29 29 55.4  &  4.32 & 4732 & 2.7 & -0.33 & \citet{mass08}  \\
       37740 & 62345   &  07 44 26.87  & +24 23 53.3  &  3.57 & 5101 & 3.04 & 0.03 & \citet{luck07} \\
       37826 & 62509   &  07 45 19.36  & +28 01 34.7  &  1.16 & 4998 & 3.13 & 0.17 & \citet{luck07} \\ 
       39424 & 66216   &  08 03 31.10  & +27 47 39.9  &  4.94 & 4560 & 2.5 & 0.03 & \citet{mass08} \\
       42527 & 73108   &  08 40 12.90  & +64 19 40.3  &  4.59 & 4564 & 2.28 & -0.16 & \citet{luck07} \\
       42911 & 74442   &  08 44 41.11  & +18 09 17.5  &  3.94 & 4763 & 2.67 & 0.01 & \citet{luck07} \\
       47029 & 82741   &  09 35 03.85  & +39 37 17.2  &  4.81 & 4934 & 2.81 & -0.12 & \citet{luck07} \\ 
       51808 & 91190   &  10 35 05.59  & +75 42 46.7  &  4.86 & 4890 & 3.07 & -0.15 & \citet{mcw90} \\
       53229 & 94264   &  10 53 18.64  & +34 12 56.0  &  3.79 & 4677 & 2.8 & -0.20 & \citet{mass08} \\
       58948 & 104979   &  12 05 12.67  & +08 43 58.2  &  4.12 & 4996 & 2.86 & -0.33 & \citet{luck07} \\
       59847 & 106714   &  12 16 20.56  & +23 56 43.5  &  4.93 & 4887 & 2.6 & -0.24 & \citet{mass08} \\
       60172 & 107328   &  12 20 21.15  & +03 18 45.8  &  4.97 & 4514 & 1.94 & -0.46 & \citet{luck07} \\
       62763 & 111812   &  12 51 41.93  & +27 32 26.6  &  4.93 & 5623 & 2.9 & 0.01 & \citet{mass08} \\
       63608 & 113226   &  13 02 10.76  & +10 57 32.8  &  2.85 & 5145 & 3.12 & 0.13 & \citet{luck07} \\
       72125 & 129972   &  14 45 14.50  & +16 57 51.9  &  4.60 & 4887 & 2.7 & -0.10 & \citet{mass08} \\
       73620 & 133165   &  15 02 54.07  & +02 05 28.6  &  4.39 & 4825 & 2.79 & -0.13 & \citet{luck07} \\
       74666 & 135722   &  15 15 30.10  & +33 18 54.4  &  3.46 & 4963 & 2.73 & -0.30 & \citet{luck07} \\
       75458 & 137759   &  15 24 55.78  & +58 57 57.7  &  3.29 & 4477 & 2.5 & 0.03 & \citet{mass08} \\ 
       77070 & 140573   &  15 44 16.00  & +06 25 31.9  &  2.63 & 4498 & 2.5 & 0.03 & \citet{mass08} \\
       77655 & 142091   &  15 51 13.94  & +35 39 29.6  &  4.79 & 4764 & 3.0 & -0.04 & \citet{mass08} \\
\enddata
\end{deluxetable}
%-----------------------------------------------------------------------

\begin{deluxetable}{lcccc}
\tabletypesize{\scriptsize}
\tablecolumns{5}
\tablewidth{0pt}
\tablecaption{Known Planet Hosts in the WES Sample }
\tablehead{
\colhead{Planet} & \colhead{Period (days)} & \colhead{M sin $i$ (\Mjup)} & 
\colhead{$a$ (AU)} & \colhead{Reference}
}
\startdata
\label{planethosts}
$\alpha$ Ari (HD\,12929) & 380.0$\pm$0.3 & 1.72$\pm$0.19 & 1.130$\pm$0.062 & \citet{lee11} \\
$\epsilon$ Tau (HD\,28305) & 595$\pm$5 & 7.7$\pm$0.3 & 1.936$\pm$0.034 & \citet{sato07} \\
$\beta$ Gem (HD\,62509) & 589.6$\pm$0.8 & 2.76$\pm$0.14 & 1.757$\pm$0.029 & \citet{hatzes93} \\
4 UMa (HD\,73108) & 269$\pm$2 & 7.1$\pm$0.6 & 0.877$\pm$0.036 & \citet{dollinger07} \\
$\iota$ Dra (HD\,137759) & 511.098$\pm$0.089 & 9.3$\pm$0.9 & 1.31$\pm$0.06 & \citet{frink02} \\
$\kappa$ CrB (HD\,142091) & 1300$\pm$15 & 1.97$\pm$0.12 & 2.72$\pm$0.05 & \citet{johnson08} \\
\enddata
\end{deluxetable}

%-----------------------------------------------------------------------

\begin{deluxetable}{lll}
\tabletypesize{\scriptsize}
\tablecolumns{3}
\tablewidth{0pt}
\tablecaption{Weihai radial velocities for HD\,62509 }
\tablehead{
\colhead{BJD-2400000} & \colhead{Velocity (\ms)} & \colhead{Uncertainty (\ms)}
}
\startdata
\label{weihaivels}
56293.24524  &     -9.0  &    7.7  \\
56293.25392  &     -6.1  &    7.9  \\
56293.26226  &     -0.3  &    7.9  \\
56293.27088  &     -7.5  &    7.4  \\
56316.09065  &    -11.9  &    8.4  \\
56316.10066  &     -2.0  &    7.9  \\
56316.11256  &    -13.7  &    8.2  \\
56316.12067  &     -2.4  &    8.6  \\
56316.12989  &    -12.9  &    8.2  \\
56317.12107  &     -6.8  &    7.2  \\
56317.13220  &     -3.0  &    7.8  \\
56320.13640  &    -16.6  &    8.0  \\
56320.14513  &    -12.8  &    8.4  \\
56320.15348  &    -11.7  &    8.1  \\
56320.16399  &    -11.6  &    8.5  \\
56320.17287  &      2.6  &    8.7  \\
56617.27237  &     73.8  &    9.2  \\
56617.28150  &     72.2  &    8.9  \\
56617.28955  &     72.6  &    9.0  \\
56617.29755  &     74.0  &    9.1  \\
56617.30750  &     83.4  &    9.1  \\
56617.31558  &     84.1  &    9.9  \\
56617.32360  &     81.1  &    8.7  \\
56617.33160  &     82.5  &    8.9  \\
56617.34040  &     79.5  &    9.4  \\
56617.34844  &     79.0  &    9.4  \\
56724.03493  &     43.2  &   10.2  \\
56749.01689  &     46.8  &    8.1  \\
56749.04009  &     48.4  &    8.7  \\
56749.07138  &     43.2  &    8.1  \\
56763.99622  &     40.0  &    7.8  \\
56764.02025  &     46.0  &    8.3  \\
56781.98304  &     27.1  &    8.6  \\
56781.99446  &     31.3  &    8.6  \\
56782.03382  &      2.5  &    8.4  \\
56786.99866  &     32.3  &    8.2  \\
56792.98783  &     13.2  &    9.5  \\
56792.99932  &     37.6  &   10.8  \\
\enddata
\end{deluxetable}

%-----------------------------------------------------------------------

\begin{deluxetable}{ll}
\tabletypesize{\scriptsize}
\tablecolumns{2}
\tablewidth{0pt}
\tablecaption{Updated Parameters for HD\,62509b }
\tablehead{
\colhead{Parameter} & \colhead{Value}
}
\startdata
\label{planetparams}
Period  & 591.2$\pm$0.76 days \\
Eccentricity & 0.071$\pm$0.028 \\
$\omega$  & 263$\pm$19 degrees \\
$K$ (\ms) & 44.1$\pm$1.2 \ms  \\
$T_0$  & 2444036.3$\pm$32.5 JD  \\
m sin $i$  & 2.80$\pm$0.08 \Mjup \\
$a$  & 1.7113$\pm$0.0015 AU  \\
\hline
RMS -- McD cs21  & 14.9\,\ms  \\
RMS -- McD phase 3  & 13.7\,\ms  \\
RMS -- TOPS  & 12.0\,\ms  \\
RMS -- CFHT  & 19.2\,\ms  \\
RMS -- DAO  & 37.2\,\ms  \\
RMS -- McD 2.1m  & 23.2\,\ms  \\
RMS -- Lick  & 9.2\,\ms  \\
RMS -- Weihai  & 7.3\,\ms  \\
\enddata
\end{deluxetable}

%-----------------------------------------------------------------------

\end{document}